\documentclass[runningheads]{llncs}
\usepackage{graphicx}
\usepackage{float}
\usepackage{csquotes}
\usepackage{tikz}
\usepackage{comment}
\usepackage{multirow}

\begin{document}
\title{Dishonesty Tendencies in Testing Scenarios Among Students with Virtual Reality and Computer-Mediated Technology}
%
%
\author{Tanja Kojić*\inst{1}\and
        Alina Dovhalevska*\inst{1}\and
        Maurizio Vergari\inst{1}\and
        Sebastian Möller\inst{1,3} \and
        Jan-Niklas Voigt-Antons\inst{2}}
\authorrunning{T. Kojić et al.}
%
\institute{Quality and Usability Lab, TU Berlin, Berlin, Germany \and Immersive Reality Lab, Hamm-Lippstadt University of Applied Sciences, Lippstadt, Germany \and German Research Center for Artificial Intelligence (DFKI), Berlin, Germany}
\maketitle              

\begin{abstract}

Virtual reality (VR) systems have the potential to be an innovation in the field of e-learning. Starting with fully functional e-classes, VR technologies can be used to build entire e-campuses. The power of VR is that it allows for stronger contact with students than computer-mediated technology. Deceptive behaviour, both verbal and nonverbal, refers to intentional activities designed to deceive others.  Students often engage in dishonest practices to make progress.

Whether it is cheating on an exam, copying another student's essay, or inflating their GPA, the motivation for cheating is rarely simply a lack of preparation. Even though some may see academic dishonesty as an asset, the reality is that it can have major consequences.

This poster demonstrates the findings from a study of students' deceitful behaviour during a test in VR and in real-life situations. For this user study, 22 volunteers were invited to test, with each experiment involving exactly two participants and the examiner present in the room. Students were invited to take two tests: one in VR and one on a laptop. Their goal was to score as many points as possible by simulating a real-world online exam. Participants were requested to complete questionnaires during and after each experiment, which assisted in collecting additional data for this study.  The results indicate that the amount of cheating that happened in VR and on a laptop was exactly the same.

\end{abstract}

\newcommand\copyrighttext{%
    \footnotesize \textcopyright 2026 IEEE. Personal use
    of this material is permitted. Permission from IEEE
    must be obtained for all other uses, in any current or
    future media, including reprinting/republishing this
    material for advertising or promotional purposes,
    creating new collective works, for resale or
    redistribution to servers or lists, or reuse of any
    copyrighted component of this work in other works.
    https://doi.org/10.1007/978-3-031-61953-3\_14}

\newcommand\copyrightnotice{%
\begin{tikzpicture}[remember picture,overlay,shift=
    {(current page.south)}]
    \node[anchor=south,yshift=10pt] at (0,0)
    {\fbox{\parbox{\dimexpr\textwidth-\fboxsep-
    \fboxrule\relax}{\copyrighttext}}};
\end{tikzpicture}%
}
\copyrightnotice
\let\thefootnote\relax\footnotetext{* Tanja Kojić and Alina Dovhalevska contributed equally to this publication.}
\setcounter{footnote}{0}

\section{Introduction}
Numerous VR research studies were focused on the technical aspect, to improve the software and hardware for better human-machine interaction and user experience. Fewer such were done to underline the possible impact of VR on already existing fields of human occupation \cite{14_VR_Walsh}.



Academia has long been using VR technologies in the process of learning. For example, VR systems are being widely used for surgical simulators, allowing medical students and doctors to practice the conduction of virtual operations \cite{14_VR_Walsh}. 
VR systems can be a potential breaking-through medium in the development of e-learning. Starting from fully operated e-classes, VR technologies can be used for creating entire e-campuses \cite{14_VR_Walsh}. The power of VR is in providing more powerful interaction with a student than computer-mediated technology. 
In dealing with human-machine interaction, human behaviour should be thoroughly analysed in the new learning environment. One of the main tasks of universities is to guarantee fair treatment for each student, as well as fair chances and possibilities on the way to gaining a degree.
%
%

Deceptive behaviours, both verbal and non-verbal, refer to intentional actions aimed at misleading others \cite{19_DePaulo2003}. Referring to Wang et al. \cite{Wang2004}, South Dakota State University scientists Hicks and Ulvestad classify that "nonverbal cues are clues of deceit that are expressed through facial expressions, eye movements, and body language. [And] verbal cues are linguistic clues of deceit that are expressed in an individual’s statement, such as stuttering, differentiation in pitch, etc." \cite{20_Hicks}. For example, a person may lie about their qualifications or experience during an interview in order to gain a job advantage over others, or a person may smile while lying to make the lie appear more believable.
The studies of deceptive behaviours \cite{4_Phillips} \cite{8_Ekman} \cite{9_Depaulo} have played a significant role in the field of psychology, providing valuable insights into the motivations behind these actions, as well as the ways in which they can be detected. Researchers \cite{2_Greene} \cite{10_Ackerman} \cite{11_Bunn} have found that deceptive behaviours are often the result of a complex interplay between individual and situational factors, and that these behaviours can have significant consequences for both the deceiver and the deceived.
It is not uncommon for students to engage in deceptive behaviours in order to get ahead. "Cheating has always been a problem in academic settings, and with advances in technology such as cell phones, and more pressure for students to score well so that they get into top-rated universities, cheating has become an epidemic" \cite{3_Batool}. 
%

\subsection{Related Work}
Much research on deceptive behaviour included experiments in which people were observed either telling a lie or telling the truth. Such studies helped to gather information about non-verbal and verbal signs of lying. Gozna, Vrij and Bull \cite{6_Gozna} compared deceptive behaviour in everyday life and in high stake scenarios. Students were put in an imaginary situation, where they were required to lie about their plagiarized essays. Additionally, participants were interviewed with the questionnaires in an attempt to understand the personal perception of deceptive behaviours under both conditions. \cite{6_Gozna} 

Greene and Saxe \cite{2_Greene} conducted research on the reasons why academic cheating behaviour appears in the first place. The goal of the study was to "explain how students, who believe that cheating is wrong, are nevertheless able to engage in cheating behaviour" \cite{2_Greene}. As to be expected, students reported more cheating done by their fellow students than by themselves. They also believed that if their peers cheat more, then they can justify their own deceptive behaviour, making it more ethically appropriate. This phenomenon was explained with 2 theories: the uniqueness bias and the theory of downward comparison. Like many other scientists, in their definition of uniqueness bias Monin and Norton refer to the work of Goethals et al. from 1991 \cite{Goethals1991}, and define it as "the tendency for people to underestimate the proportion of others who can or will perform desirable actions [...]. In practice, those who perform a desirable behaviour underestimate the number of others as good as them, whereas those who perform an undesirable behaviour overestimate the number of others as bad as them" \cite{uniqueness_bias_def}. Referring to a work of Wills from 1991 \cite{Wills1991}, Greene and Saxe explained the theory of downward comparison as "people will compare themselves to others who are worse off than they are, in order to appear better themselves" \cite{2_Greene}. Moreover, it was established during this study that students thought their classmates benefit more from cheating. Meaning that they personally would get the same result without cheating, but others needed cheating to pass. This allowed them to again rationalize their deceptive behaviour. \cite{2_Greene}


University exams are inevitably connected with stress in students' life. And since "human beings are radically social by nature" \cite{21_bailey}, one of the best ways to cope with stress is having friends. 

Human behaviour in VR and real life has been widely compared in many experimental studies and research papers. For example, {\"O}zg{\"u}r G{\"u}rerk and Alina Kasulke \cite{example_paper1} studied the difference in charitable behaviour in real and virtual environments, Rajaram Bhagavathula et al. \cite{example_paper2} $-$ compared pedestrian behaviour.

Various research studies were done to find out how students cheat and why they cheat. Furthermore, the targeted environment to detect deception was either virtual reality or real-world circumstances. To the best of our knowledge, there is no research that compares students' cheating behaviour using both VR and computer-mediated technology. And while other researchers focused more on having a virtual observer in their studies, the current study was conducted under conditions of a real observer being present in the room under both conditions.

\subsection{Research questions}
The goal of the present study was to compare the possible deceptive behaviour of students using VR and computer-mediated technology. Through our experiment, we seek to answer whether virtual reality would have an impact on students' deceptive behaviour during exams.

\textbf{RQ1:} Will students cheat more or feel the urge to cheat more during a test in virtual reality, than while doing it on a laptop? 

\textbf{RQ2:} Will a pair of participants, who were familiar with each other, feel more relaxed and at ease to cheat during a test? 

\textbf{RQ3:} And on the contrary, will a pair of participants, who were not familiar with each other, cheat neither in VR nor on a laptop?

\section{Method}
\label{sec:Methodology}
"There are, in principle, three ways to catch liars: (1) by observing how they behave (the movements they make, whether they smile or show gaze aversion, their pitch of voice, their speech rate, whether they stutter, and so on), (2) by listening to what they say (analysing the speech content), and (3) by measuring their physiological responses." \cite{1_Vrij}

\subsection{Data collection}
The participants were invited to the Quality and Usability Lab on a campus of TU Berlin. Each experiment was held with exactly 2 participants and the researcher present in the room. Students were asked to take 2 tests - one in VR and another on a laptop. Their goal was to gain as many points as possible by putting themselves into the situation of a real online exam.
In order to gather data, the history of the VR browser and the laptop browser were checked after each participant, and the screens on both laptops were recorded as well. 

\subsection{Participants}
Participants were 22 students from respective Berlin universities, with an average mean of 11 semesters of study time. Among them, 14 males with a mean age of 26 and 7 females with a mean age of 26. One participant preferred not to specify gender. On a scale from 1 to 5, participants' average VR experience score was 2. All students had participated in online exams before and estimated their level of anxiety during an exam as 3 out of 5. 

\subsection{Study setup}
On average, one experiment took around 40 minutes.
After reading an information sheet about the present study, students were asked to sign a consent form and fill out the Demographics. Both first and second tests were defined as conditions of "Laptop" and "VR". To minimize order-effects \cite{order_effects} on the results of the study, conditions were randomized using the Latin squares system \cite{latin_squares}. Before each test, a short instruction was given in which it was explained how participants can navigate through the test, how much time they have, and in the case of VR, how to use a controller. The examples of how participants had seen the tests on a laptop and in VR are illustrated in Figures 1 and 2. 

\begin{figure}[!ht]
\includegraphics[width=\textwidth]{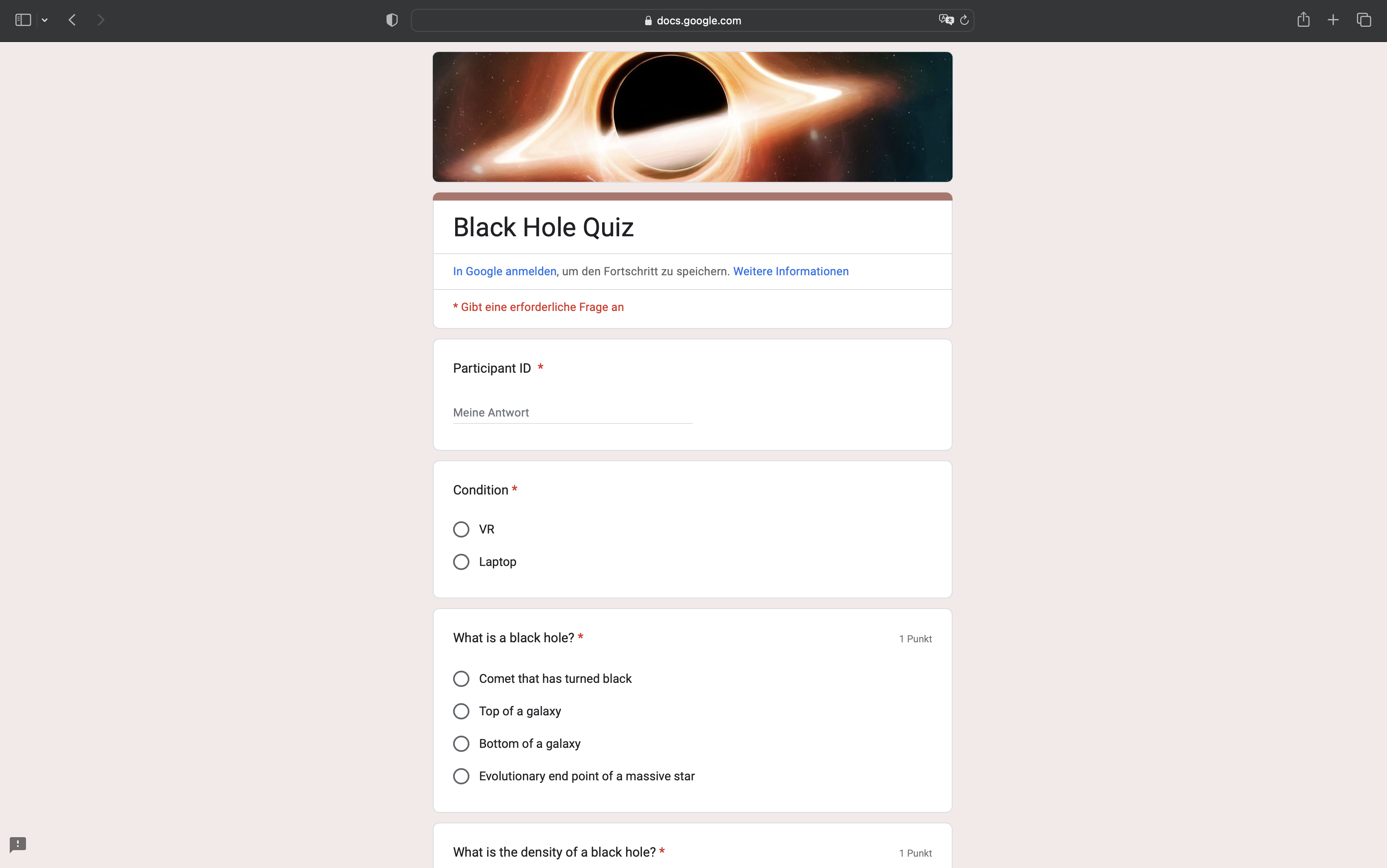}
\caption{Example of a quiz "Black Hole", which participants should have done on a laptop}
\label{fig:ui_laptop}
\vspace{-1em}
\end{figure}

\begin{figure}[!ht]
\includegraphics[width=\textwidth]{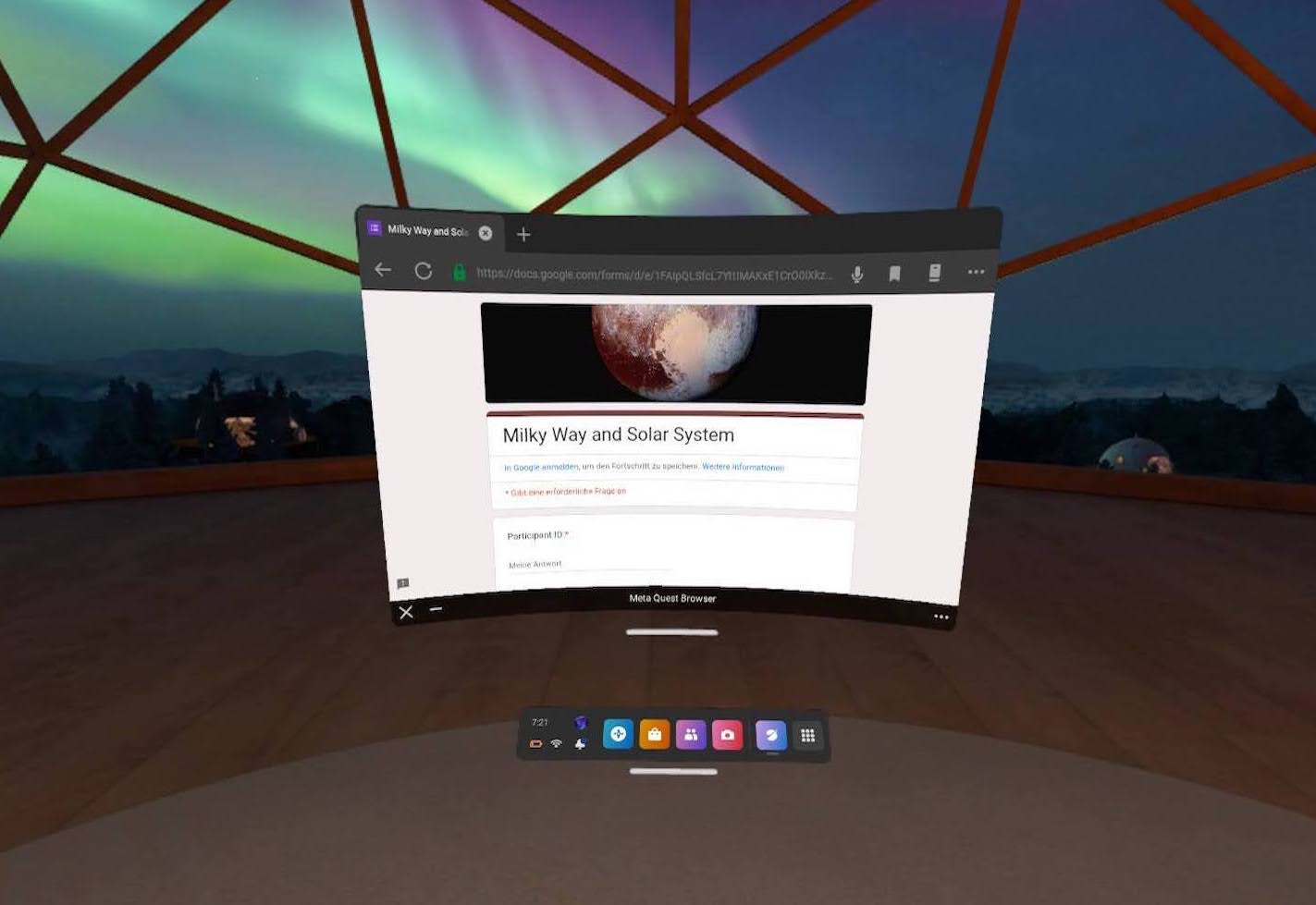}
\caption{Example of a quiz "Milky Way and Solar System", which participants should have done in VR}
\label{fig:ui_vr}
\vspace{-1em}
\end{figure}

After each test (in VR and on a laptop) participants filled out the SAM \cite{sam} and IPQ \cite{ipq} forms and evaluated how they were feeling during the experiment and after it. 
The students were asked to fill out the questionnaires "Self-perception of Lying" and "Perception of Others Lying" at the end of the experiment after the study goal was revealed to them. Hence, they had an opportunity to reflect on their feelings and thoughts during the experiment to describe them.

\section{Results and Discussion}
\label{sec:Results}

The data from the SAM and IPQ questionnaires were analyzed in order to find the significant differences between the conditions of "Laptop" and "VR". Therefore, independent samples t-tests were run with open statistical software "jamovi" \cite{jamovi}. As a result, conditions of "VR" and "Laptop" were significantly different in the following questions:

\begin{table}[h]
\caption{Independent Samples T-Test}
\centering
\begin{tabular}{|p{7.5cm}|c|c|c|}
\hline Question & Statistic & df & $\mathrm{p}$ \\
\hline 
In the computer generated world I had a sense of "being there" & $-3.727^a$ & 42.0 & $<.001$ \\
\hline 
Somehow I felt that the virtual world surrounded me. & -6.653 & 42.0 & $<.001$ \\
\hline 
I did not feel present in the virtual space. & -4.723 & 42.0 & $<.001$ \\
\hline 
I had a sense of acting in the virtual space, rather than operating something from outside. & -7.456 & 42.0 & $<.001$ \\
\hline 
I felt present in the virtual space. & -6.452 & 42.0 & $<.001$ \\
\hline 
How aware were you of the real world surrounding while navigating in the virtual world (i.e. sounds, room temperature, other people, etc.)? & -4.313 & 42.0 & $<.001$ \\
\hline 
I was not aware of my real environment. & -3.389 & 42.0 & 0.002 \\
\hline 
I still paid attention to the real environment. & 3.900 & 42.0 & $<.001$ \\
\hline 
I was completely captivated by the virtual world. & -4.688 & 42.0 & $<.001$ \\
\hline 
How real did the virtual world seem to you? & -2.451 & 42.0 & 0.018 \\
\hline
\end{tabular} 
\end{table}

\begin{table}[h]
\caption{Group Descriptives}
\centering
\begin{tabular}{|p{7cm}|c|c|c|}
\hline Question & Condition & Mean & SD \\
\hline \multirow[t]{2}{=}{In the computer generated world I had a sense of being there} & Laptop & 2.18 & 1.532 \\
\cline{2-4} & VR & 3.64 & 1.002 \\
\hline \multirow[t]{2}{=}{Somehow I felt that the virtual world surrounded me.} & Laptop & 1.73 & 1.032 \\
\cline{2-4} & VR & 3.82 & 1.053 \\
\hline \multirow[t]{2}{=}{I did not feel present in the virtual space.} & Laptop & 1.73 & 1.032 \\
\cline{2-4} & VR & 3.36 & 1.255 \\
\hline \multirow[t]{2}{=}{I had a sense of acting in the virtual space, rather than oper} & Laptop & 1.55 & 0.800 \\
\cline{2-4} & VR & 3.59 & 1.008 \\
\hline \multirow[t]{2}{=}{I felt present in the virtual space.} & Laptop & 1.86 & 1.037 \\
\cline{2-4} & VR & 3.77 & 0.922 \\
\hline \multirow[t]{2}{=}{How aware were you of the real world surrounding while navigat} & Laptop & 2.09 & 1.065 \\
\cline{2-4} & VR & 3.50 & 1.102 \\
\hline \multirow[t]{2}{=}{I was not aware of my real environment.} & Laptop & 2.09 & 1.231 \\
\cline{2-4} & VR & 3.27 & 1.077 \\
\hline \multirow[t]{2}{=}{I still paid attention to the real environment.} & Laptop & 3.82 & 1.259 \\
\cline{2-4} & VR & 2.50 & 0.964 \\
\hline \multirow[t]{2}{=}{I was completely captivated by the virtual world.} & Laptop & 1.64 & 1.049 \\
\cline{2-4} & VR & 3.00 & 0.873 \\
\hline \multirow[t]{2}{=}{How real did the virtual world seem to you?} & Laptop & 1.95 & 1.327 \\
\cline{2-4} & VR & 2.86 & 1.125 \\
\hline
\end{tabular}
\end{table}

Present research demonstrated that cheating occurred among 5 pairs of participants who were familiar with each other. Among them: 
\begin{itemize}
    \item 2 students who cheated only in VR;
    \item 2 students who cheated only on a laptop;
    \item 6 students who cheated under both conditions.  
\end{itemize}

Contrary to the previous data, the other 5 pairs of participants were as well familiar with each other but did not cheat. This result could be supported by the answers received in questionnaires, which indicated a high percentage of participants feeling guilt after deception. Further findings indicated 1 pair of students who were not familiar with each other and likewise did not cheat. 

Within 11 subjects, who did not cheat but had an urge to, were 3 students who thought about it in VR, and 5 - who thought about it while doing tests both in VR and on a laptop. They reported several reasons why they did not cheat. Such as the fact that this would indicate their lack of intelligence, or simply because they were observed. Mol stated that "the observer triggers feelings of being watched, which activates a concern for reputation and a desire to abide by the norm of honesty" \cite{17_Mol}, And the majority of the students were worried about being caught in deception.

\section{Conclusion}
\label{sec:Conclusion} 

Considering the rapid development and expansion of VR innovations, it is beneficial for researchers to focus on having VR technologies in the educational paradigm. In 2002 Walsh wrote that "with a strong theoretical foundation established, high quality experiments can be designed which measure the effects of VR systems. It may be appropriate to test some aspects of VR systems with captive audiences such as university students, while other aspects will need study in the richness of real world environments" \cite{15_VR_Walsh}. Our study is among others to contribute to a better understanding of students' deceptive behaviour in VR compared to real life. 

Certain limitations were met during this research study. Due to ethical reasons, we had to inform the participants in the consent form about screen recording. It could potentially influence the behaviour of subjects during tests. 
A part of the participants of this study was familiar with the researcher since they studied at the same university. This also means that they were not feeling as intimidated by the researcher as they would be by a professor or teaching assistant. Moreover, since it was not a real exam, there was no feeling of being put under pressure. 
Future research directions for this paper include investigating the effects of a researcher's presence on participant behavior in virtual reality, examining the impact of VR proficiency on study results, and exploring psychological group dynamics by involving more than two subjects simultaneously.

\bibliographystyle{splncs04}
\bibliography{main}
\end{document}